\documentclass[pre,twocolumn,floatfix]{revtex4-2}
\pdfoutput=1 
\usepackage{graphicx}
\usepackage{graphics}
\usepackage{latexsym}
\usepackage{amsmath, amsthm, amssymb, wasysym}
\usepackage{xcolor}
\usepackage{epsfig}
\usepackage{soul}
\usepackage{cancel}
\usepackage{hyperref}
\usepackage{orcidlink}

\begin{document}
\title{ Synergy of Doob Transformation and Montroll Defect Theory for Random Walks in External Potentials 
}

\author{Stanislav Burov$^{1}$\orcidlink{0000-0003-1065-174X}}

\email{stasbur@gmail.com}
\affiliation{$^1$Department of Physics, Bar-Ilan University, Ramat-Gan 5290002,
Israel}



\begin{abstract}

We present a systematic method for constructing stochastic processes by modifying simpler, analytically solvable random walks on discrete lattices. Our framework integrates the Doob $h$-transformation with the Montroll defect theory, overcoming the strict constraints associated with each method alone. By combining these two approaches, we map random walks in simple potentials onto processes involving more general external potentials and metastable states. Explicit analytical expressions relate the transformed process to the original one, facilitating direct investigation of exponential decay rates and additional dynamical modes. 
As an illustrative example, we demonstrate our method by analyzing a random walker in a linear potential modified to include a metastable state, revealing distinct exponential decay regimes relevant to escape dynamics. 
\end{abstract}

\maketitle

\newcommand{\fr}{\frac}
\newcommand{\lr}{\langle}
\newcommand{\rl}{\rangle}
\newcommand{\tl}{\tilde}


\section{Introduction }

When a particle undergoes random motion in the presence of a confining or driving potential, its probability distribution function (PDF) is shaped by the competition between stochastic fluctuations and deterministic forces. 
In many cases, however, obtaining the PDF in the presence of a complicated external potential is challenging and typically requires numerical simulations or asymptotic approximations~\cite{risken1996fokker}. A central question is whether such complex stochastic dynamics can be effectively represented using simpler, analytically tractable systems.

In this study, we develop a systematic framework for constructing complex stochastic processes by modifying simpler, analytically solvable ones. 
We focus on random walks on discrete lattices with general transition probabilities, combining the Doob $h$-transformation~\cite{doob1984classical} with the inclusion of branching and defects~\cite{montroll1964random,hughes1996random} at selected sites. 
While the Doob $h$-transformation introduces effective potentials, it also imposes strict constraints. 
The addition of branching and defects eases these constraints, ensuring the method’s broader applicability.
By bringing both elements together, we can express the probability of the transformed process directly in terms of the original system, thus linking a basic, tractable process to a more intricate one. This synergy provides analytical insights and a practical route to modeling non-trivial dynamical behaviors. 
In particular, it allows us to examine the decay-mode spectrum of the modified process by analyzing the generating functions of the original model.

As an illustrative example, we  consider a discrete random walker in a linear potential and show how modifying transition probabilities and introducing controlled branching at two lattice sites, maps the process onto motion in a potential with a metastable state and potential barrier of height $\Delta U$. 
Using the Doob $h$-transformation, we explicitly express the z-transform of the probability of the modified process in terms of the original one, establishing a direct analytical connection. 
Specifically, we find the exponential decay rate of the probability function and its dependence on the Kramers escape rate $\exp(-\Delta U/k_B T)$ not only in the large $\Delta U$ limit but also for situations when the potential barrier is wide and $\Delta U$ is small.     

The manuscript is structured as follows: In Sec.\ref{sec:Map}, we derive the mapping between a random walk in a simple potential and a random walk in a more complex one. 
In Sec.\ref{sec:metaexample}, we apply this method explicitly to explore a potential exhibiting a metastable state. 
Finally, Sec.~\ref{sec:discussion} provides a summary and discusses the implications of our results.

\section{Construction of a General Mapping}
\label{sec:Map}

 Consider a discrete Markov process on a lattice with one-step transition probabilities $W(y | x)$, which describe the probability of transitioning from site $x$ to site $y$. The evolution of the probability distribution is governed by
\begin{equation}
P_{n+1}(y| x_0) = \sum_{x} W(y| x)\, P_n(x| x_0),
\end{equation}
$P_n(y| x_0)$ denotes the probability to reach $y$ in $n$ steps starting from $x_0$.
The Doob $h$-transformation~\cite{doob1984classical} reweights the process by introducing a positive function $h(x)$, so that the transformed transition probabilities are defined as
\begin{equation}
W^*(y | x) = \frac{h(y)}{h(x)}\,W(y | x).
\end{equation}
A key property of this transformation is the telescopic cancellation along any trajectory. Specifically, for a trajectory
\begin{equation*}
x_0 \to x_1 \to \cdots \to x_n,
\end{equation*}
the cumulative weight factor is
\begin{equation*}
\frac{h(x_1)}{h(x_0)} \cdot \frac{h(x_2)}{h(x_1)} \cdots \frac{h(x_n)}{h(x_{n-1})} = \frac{h(x_n)}{h(x_0)}.
\end{equation*}
Thus, the Doob $h$-transformation affects identically any two trajectories that start at $x_0$ and end at $x_n$. 
In particular, after the Doob $h$-transformation the probability of the transformed process is
\begin{equation}
P_n^*(y| x_0)=\frac{h(y)}{h(x_0)}\,P_n(y| x_0).
\label{eq:doobproababilitytransform}
\end{equation}
Equation~\eqref{eq:doobproababilitytransform} shows how the probability structure of a process with a given set of transition rules can be mapped onto a new process with a different, potentially more intricate set of transitions, while preserving the statistical properties of trajectories.

For the transformed process to define a valid probability distribution, the modified transition probabilities must sum to one. This imposes the harmonicity condition on $h(x)$,
\begin{equation}
\label{eq:harmonicity}
\sum_y W(y| x)\, \frac{h(y)}{h(x)} = 1,
\end{equation}
which ensures that the reweighted probabilities remain normalized.
The harmonicity condition in Eq.~\eqref{eq:harmonicity} essentially states that $h(x)$ is an eigenfunction of the transition operator $W$ with eigenvalue 1, i.e., 
$\sum_y W(y| x)\, h(y) = h(x)$
.
This requirement imposes harsh constraints on $h(x)$. 
It restricts $h(x)$ to a very specific set of functions—namely, those that are harmonic with respect to the dynamics of the original process. In many cases, Eq.~\eqref{eq:harmonicity}  has a unique solution (up to an overall multiplicative constant), leaving little room for arbitrary choices. 
If $h(x)$ does not satisfy the harmonicity condition, the reweighted transition probabilities would not sum to one, leading either to a loss or an excess of probability, and consequently to a process that is not properly normalized.
Therefore, the harmonicity constraint tightly restricts the admissible forms of $h(x)$ and ensures that only those reweightings that leave the probability flow invariant are permitted.

Instead of enforcing the harmonicity condition by adjusting $h(x)$, we take an alternative approach: we allow $h(x)$ to be arbitrary and instead modify the original process by introducing additional branching, effectively reweighting the transition probabilities $W(y| x)$ to a new set $\tilde{W}(y| x)$ such that the modified process satisfies Eq.~\eqref{eq:harmonicity}.
To achieve this, we define $\tilde{W}(y| x)$ as a modification of the original transition probabilities by adding a correction term at $x$:
\begin{equation}
\tilde{W}(y| x) = W(y| x) + \Delta(y| x).
\label{eq:wtildedefinition}
\end{equation}
Where $\Delta(y| x)$ 
represents an additional probability mass (or, if negative, a deficit) directly assigned to the transition from $x$ to $y$. Although these reweighted probabilities do not necessarily sum to one, the critical requirement is that the function $h(x)$ becomes harmonic with respect to $\tilde{W}$; that is, we must have
\begin{equation}
\sum_y \tilde{W}(y| x)\, h(y) = h(x),
\label{eq:wtildeharmonicity}
\end{equation}
which ensures that the transformed process remains properly normalized. Substituting Eq.~\eqref{eq:wtildedefinition} into Eq.~\eqref{eq:wtildeharmonicity} gives
\begin{equation}
\sum_y \Delta(y| x)\, h(y) = h(x) - \sum_y W(y| x)\, h(y).
\end{equation}
which determines the total $h$-weighted probability mass to be added or removed at state $x$, though it does not uniquely fix the individual terms $\Delta(y| x)$.

Thus, instead of requiring $h(x)$ to be an eigenfunction of $W(y|x)$, we reinterpret deviations from harmonicity as additional branching or killing events at position $x$, encoded by the function $\Delta(x'|x)$. Consequently, the Doob $h$-transformation does not directly map $P_n(y|x_0)$ to $P_n^*(y|x_0)$; rather, we first construct an intermediate reweighted process with probability ${\widetilde P}_n(y|x_0)$, obtained by mapping the original kernel $W$ onto a new kernel ${\widetilde W}$, and only then map ${\widetilde P}_n$ to $P_n^*$ via the Doob-$h$ transformation.

We now provide the explicit renewal equation linking $P_n(y | x_0)$ and ${\tilde P}_n(y | x_0)$. The reweighting occurs only at defect sites $x$ belonging to a set $B$.  Using the Montroll approach for defect-site problems~\cite{montroll1964random,hughes1996random}, we have
\begin{equation}
\begin{array}{l}
{\tilde P}_n(y | x_0) 
= P_n(y | x_0) 
+ 
\\
\sum_{m=0}^{n-1}\sum_{x\in B} P_m(x | x_0)\sum_{x'}\Delta(x' | z)\,{\tilde P}_{\,n-m-1}(y| x')
\end{array}
\label{eq:renewal}
\end{equation}
The first term in Eq.~\eqref{eq:renewal} represents the original, unperturbed dynamics, while the second (renewal) term accounts for branching (positive) or killing (negative) corrections at defect sites, summing over all possible arrival events at these sites.

Taking the $z$-transform (generating function) of the renewal equation explicitly connects the original and reweighted processes. Define the generating functions $G_z(y|x_0)=\sum_{n=0}^{\infty} z^n P_n(y|x_0)$ and $\widetilde{G}z(y|x_0)=\sum_{n=0}^{\infty} z^n \widetilde{P}_n(y|x_0)$. By reindexing the sums with $n = m + 1 + k$, we rewrite
$
\sum_{n=0}^{\infty}
\sum_{m=0}^{n-1}
P_m(x| x_0)\,\widetilde{P}_{n-m-1}(y| x{\prime})\,z^n=
\sum_{m=0}^{\infty}
\sum_{k=0}^{\infty}
P_m(x| x_0)\,\widetilde{P}_{k}(y| x{\prime})\,z^{m+1+k}
$. 
Therefore the $z$-transform of Eq.~\eqref{eq:renewal} yields
\begin{equation}
\label{eq:renewalZtranform}
    \begin{array}{l}
\widetilde{G}_z(y | x_0)
=
\\
G_z(y | x_0)
+
z \sum_{x \in B} G_z(x | x_0)\;
\sum_{x{\prime}} \Delta(x{\prime} | x)\,\widetilde{G}_z(y | x{\prime}).
\end{array}
\end{equation}
We define $A$ as the set of all sites $x'$ for which $\Delta(x'|x)\neq0$ with $x\in B$. Then, define vectors $\vec{\mathbf{f}}(y)$ and $\vec{\mathbf{g}}(y)$ whose components are $\widetilde{G}_z(y|x')$ and $G_z(y|x')$, respectively, with $x'\in A$. 
Eq.\eqref{eq:renewalZtranform} thus becomes $\vec{\mathbf{f}}(y)=\vec{\mathbf{g}}(y)+z \mathbf{M}\vec{\mathbf{f}}(y)$,
where the matrix $\mathbf{M}$ has entries,  $M_{x'',x'}=\sum_{x\in B} G_z(x|x'')\Delta(x'|x)$, while $x'',x'\in A$. 
Therefore, $\vec{\mathbf{f}}=\bigl(\mathbf{I}-z\mathbf{M}\bigr)^{-1}\vec{\mathbf{g}}$, where $\mathbf{I}$ is the identity matrix.
Let $\vec{\mathbf{w}}(x_0)=\sum_{x\in B} G_z(x|x_0)\Delta(x'|x)$ with $x'\in A$,
then
\begin{equation}
    \label{eq:gtildeZfinal form}
\widetilde{G}_z\bigl(y | x_0\bigr)
\;=\;
G_z(y|x_0)+
z\vec{\mathbf{w}}(x_0)\bigl(\mathbf{I} - z\,\mathbf{M}\bigr)^{-1}\,\vec{\mathbf{g}}(y),  
\end{equation}
and the generating function of $P^*_n(y| x_0)$ is  
\begin{equation}
    G^*_z(y| x_0) = \frac{h(y)}{h(x_0)}\widetilde{G}_z\bigl(y | x_0\bigr)
    \label{eq:gstarGeneral}
\end{equation}
while $P^*_n(y|x_0)$ is found by standard methods of $z$-transform inversion.
Equations \eqref{eq:gtildeZfinal form} and \eqref{eq:gstarGeneral} together provide a complete description of the transformation from the original process that is defined by $P_n$ to the transformed process define by $P_n^*$.  

Notice that the exponential decay of $P_n^*(y|x_0)$ with $n$ is determined solely by the poles in $z$-space, corresponding to the decay exponents $E_{\alpha}$ appearing in the expansion $P_n^*(y|x_0)=\sum_{\alpha}a_{\alpha}(n)e^{-E_{\alpha} n}$. Due to the structure of Eqs.~(\ref{eq:gtildeZfinal form}-\ref{eq:gstarGeneral}) and the fact that vectors $\vec{\mathbf{f}}$ and $\vec{\mathbf{g}}$ are constructed from the generating functions $G_z(x'|x)$ of the original process, new poles in the transformed generating function $G_z^*(y|x)$ emerge exclusively from the factor $(\mathbf{I}-z\mathbf{M})^{-1}$. Consequently, the new eigenvalues $E_{\alpha}$, defining the decay exponents, are solutions of $\det(\mathbf{I}-z\mathbf{M})=0$. We will explicitly illustrate this connection through an example involving a metastable state.

The presented mapping combines two key techniques: the Doob $h$-transformation and the Montroll renewal approach for defective sites. Each has its limitations: the $h$-transformation imposes a strict harmonicity requirement on $h(x)$, while the Montroll approach can handle any number of local transition-rule modifications but grows increasingly complex as the set of defect sites expands (since the dimension of $\mathbf{M}$ in the inverse matrix term also expands). By mixing these methods, one can overcome the harmonicity constraint at only a handful of sites and apply local corrections there, without resorting to a large-scale Montroll defect technique treatment. 
In this way, most transition probabilities are altered via the simpler $h$-transformation, keeping the size of $\mathbf{M}$ manageable. 
We illustrate this synergy next with a concrete example of motion in an external potential featuring a metastable state.


\begin{figure}[t]
 \centering
 \includegraphics[width=0.90\linewidth,trim=145 470 110 60, clip]{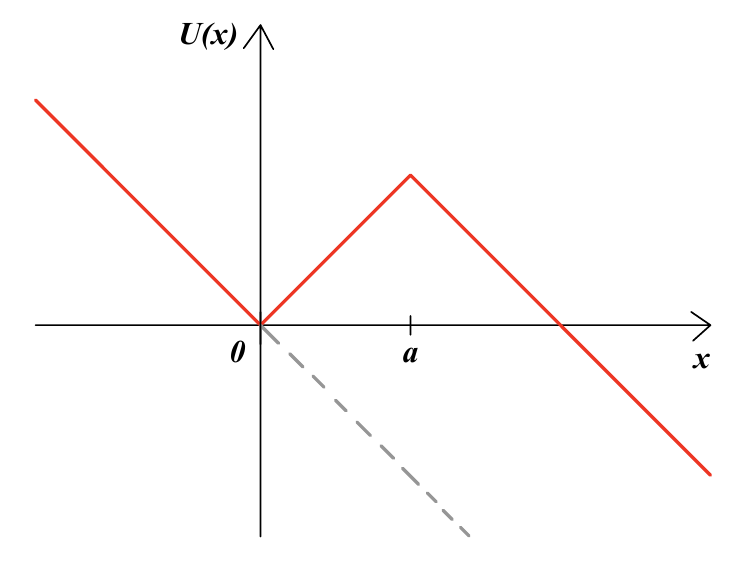}  
 \caption{
 The external ``zigzag" potential $U(x)$ with a metastable region, minima at $x=0$ and maxima at $x=a$ (red line), is obtained by Doob $h$-transformation of linear potential (dashed line) that represents motion with a presence of constant driving force. 
 The harmonicity of the transformation is broken in the bending points of $U(x)$, i.e., $x=0$ and $x=a$.
}\label{fig:zigzagpotential}
\end{figure}


\section{Example: Potential with a metastable state}
\label{sec:metaexample}

We now illustrate how the Doob $h$-transformation, combined with Montroll corrections, can create a metastable state in a one-dimensional biased random walk. The original process evolves on a one-dimensional lattice with unit spacing, restricted to nearest-neighbor transitions. Specifically,
$
W(x'|x) \;=\; q\,\delta_{x',x+1} \;+\; p\,\delta_{x',x-1}
$,
where $\delta_{x',x}$ is the Kronecker delta, $q>p$, and $p+q=1$. 
Our goal is to modify this process so that, in the region $0 \le x \le a$, the direction of the external force is reversed, thereby inducing a local potential barrier (Fig.~\ref{fig:zigzagpotential}).

Concretely, for $x<0$ or $x>a$, we keep $W^*(x'|x) = W(x'|x)$; but for $0 < x < a$, we set
$W^*(x'|x) = p\,\delta_{x',x+1} \;+\; q\,\delta_{x',x-1}$. 
This can be achieved via the following $h$-function:
\begin{equation}
    \label{eq:hfunctexample}
    h(x) =
\begin{cases} 
1, & x < 0 \\[8pt]
\left(\frac{p}{q}\right)^x, & 0 \leq x \leq a \\[8pt]
\left(\frac{p}{q}\right)^a, & x > a
\end{cases}
\end{equation}
One readily verifies that $W(x'|x)\,\tfrac{h(x')}{h(x)} = W^*(x'|x)$ everywhere except at $x=0$ and $x=a$, where the sum of transformed transition probabilities falls short of unity or exceeds unity, respectively. Consequently, to restore normalization, we introduce branching/killing corrections at $x=0$ and $x=a$ via Eq.~\eqref{eq:wtildedefinition}.

{\it Branching/killing corrections.}
At $x=0$ we can conserve the symmetry of jumping left or right by introducing a new branching probability $\Delta(0|0) = 1-2p$, so
$\widetilde{W}(0|0) = 1-2p$ and,$\sum_{x'}\widetilde{W}\bigl(x'|0\bigr)\,\tfrac{h(x')}{h(0)} = 1$.

At $x=a$, we must reduce the overly large transformed rate $2q$ by introducing negative probability mass, i.e., ``killing". 
Since we don’t want negative probability mass to appear
after Doob $h$- transformation, we will modify the left jump
probability:
$
\Delta(a-1|a) = -\,p\,\frac{(q-p)}{q}$ and $\sum_{x'}\widetilde{W}\bigl(x'|a\bigr)\,\tfrac{h(x')}{h(a)}=1$. 
Hence, Eq.~\eqref{eq:renewalZtranform} for the generating function $\widetilde{G}_z(y|x_0)$ becomes
\begin{equation}
    \label{eq:exmplzrenewal}
    \begin{array}{ll}
    \widetilde{G}_z(y|x_0) =
    &
    G_z(y|x_0) + 
    z G_z(0|x_0)\Delta(0|0)
    \widetilde{G}_z(y|0)
    \\
    &
    +
    z G_z(a|x_0)\Delta(a-1|a)
    \widetilde{G}_z(y|a-1).
    \end{array}
\end{equation}
Comparing with Eq.~\eqref{eq:renewalZtranform}, we identify
$\vec{\mathbf{w}}(x_0)
\;=\;
(1-2p)\Bigl[G_z(0|x_0)\;,\;-\,\tfrac{p}{1-p}\,G_z(a|x_0)\Bigr]$, and 
$\vec{\mathbf{g}}^T
=
\bigl[G_z(y|0),\;G_z(y|a-1)\bigr]$,

and
\begin{equation}
\mathbf{M}
\;=\;
(1-2p)\,
\begin{pmatrix}
G_z(0|0) & -\,\tfrac{p}{1-p}\,G_z(a|0) \\[6pt]
G_z(0|\,a-1) & -\,\tfrac{p}{1-p}\,G_z(a|\,a-1)
\end{pmatrix}.
\label{eq:exmplMatrix}
\end{equation}
Substituting into Eq.~\eqref{eq:gstarGeneral} yields $G_z^*(y|x_0)$. Crucially, $\mathbf{M}$ is only $2\times2$, irrespective of $a$. In contrast, an approach based purely on the Montroll defect theory (without the Doob $h$-transformation) would require modifying every site between $0$ and $a$, producing a matrix of size $2a\times 2a$.

{\it Decay spectrum.}
To analyze the long-time decay, we use the fact  that the generating function $G_z(y|x)$ for the biased binomial random walk is~\cite{hughes1996random}
\begin{equation}
    \label{eq:exmplfreegenfunc}
    \begin{array}{l}
G_z(y|x_0)=
\\
\frac{1}{\sqrt{1-4p(1-p)z^2}}\bigl(\frac{1-\sqrt{1-4p(1-p)z^2}}{2\sqrt{p(1-p)}z}\bigr)^{|y-x_0|}\bigl(\sqrt{\frac{1-p}{p}}\bigr)^{y-x_0}.
\end{array}
\end{equation}
 The generating function 
$G_z(y|x)$ has a single pole at $z_1=1/(2\sqrt{p\,q})$, implying an exponential decay rate $\lim_{n\to\infty}\left(-\log(P_n(y|x_0))/n\right)=\ln z_1$. The ``zigzag'' potential in Fig.~\ref{fig:zigzagpotential}  introduces additional poles through the term $\det\bigl(\mathbf{I}-z\,\mathbf{M}\bigr)=0$ in Eq.~\eqref{eq:gtildeZfinal form}. Writing $z=1+w$, we obtain an equation for $w$ from
$\det\bigl(\mathbf{I}-(1+w)\,\mathbf{M}\bigr)=0
$, 
\begin{widetext}
\begin{equation}
    \label{eq:exmpldetzero}
      w
\frac{3+2w+\cosh\left(\frac{\Delta U}{a}\right)+\sinh\left(\frac{\Delta U}{a}\right)\sqrt{1-\left(\frac{1+w}{\cosh\bigl(\frac{\Delta U}{2a}\bigr)}\right)^2}}{2\left(\frac{1+w}{\cosh\left(\frac{\Delta U}{2a}\right)}\right)^2\cosh\left(\frac{\Delta U}{2a}\right)\sinh\left(\frac{\Delta U}{2a}\right)^2}
- \left(\frac{1-\sqrt{1-\left(\frac{1+w}{\cosh\bigl(\frac{\Delta U}{2a}\bigr)}\right)^2}}{\frac{1+w}{\cosh\bigl(\frac{\Delta U}{2a}\bigr)}}\right)^{2a-1}
=0
\end{equation}
\end{widetext}
Here, $\Delta U = U(a) - U(0)$. We use detailed balance for jumps in the region $0 < x < a$ to relate $U(x)$ and $p$:
$\exp\bigl[-U(x)\bigr] \, p 
=
\exp\bigl[-U(x+1)\bigr]\,\bigl(1-p\bigr)
$, which implies
$
p = \frac12 \;-\; \frac{1 - e^{-\Delta U/a}}{2\bigl(1 + e^{-\Delta U/a}\bigr)}
$
We set $k_B T = 1$ so that $\Delta U/k_B T = \Delta U$. Since the only pole of the unperturbed generating function $G_z(y|x)$ is at $z_1 = \cosh\bigl(\tfrac{\Delta U}{2a}\bigr)$, we seek solutions $z_2 = 1 + w < \cosh\bigl(\tfrac{\Delta U}{2a}\bigr)$. Introducing a metastable state slows the exponential decay of $P_n^*(y|x)$, thus adding at least one new pole closer to $1$ than $z_1$.

Equation~\eqref{eq:exmpldetzero} reveals that the quantity $w$ depends on both the height and the width of the barrier, i.e.\ $w(\Delta U/a, a)$. Consequently, there are two  limiting cases to consider:
(i) $\Delta U/a \to \infty$, which corresponds to a very \emph{high} (yet relatively narrow) barrier. This limit is closely related to the classical Kramers escape problem~\cite{hanggi1990reaction}. (ii) 
$a \to \infty$, which corresponds to a \emph{wide} barrier, even if $\Delta U$ is large. 
We set $k_BT=1$, and $\Delta U$ must be compared not only to $1$ but also to $a$. Thus ``$\Delta U$ is large" implies $\Delta U \gg1$, yet 
we still can obtain two limits that would depend if $\Delta U\to \infty$ or $a\to\infty$ is taken first.

{\it Case (i): $\Delta U /a\to \infty$.}
In this scenario, Eq.~\eqref{eq:exmpldetzero} simplifies to
\begin{equation}
    \label{eq:exmplLargeUlimit}
w\!\bigl(\tfrac{\Delta U}{a}, a\bigr) 
\;\longrightarrow\;
e^{-\Delta U}
\quad
\text{as}
\quad
\Delta U/a\to\infty.
\end{equation}
Physically, $w$ becomes the Kramers rate factor $e^{-\Delta U}$, reflecting an Arrhenius-like barrier crossing probability. Since $1 + w < \cosh\bigl(\tfrac{\Delta U}{2a}\bigr)$, the dominant pole in $z$-space has magnitude $z_2 = 1 + w$ and the long-time decay is set by
$-\lim_{n\to\infty}\left(\log P_n^*(y|x)/n\right)
=
\log(1 + w)\sim w$.
Hence, in the limit $\Delta U/a\to\infty$, one obtains
\begin{equation}
\label{eq:exmplpnlargeU}
P_n^*(y|x)
\;\sim\;
\exp\!\bigl(-\,e^{-\Delta U}\,n\bigr)
\quad
\text{as}
\quad
\Delta U/a\to\infty
.
\end{equation}
This exponential decay rate is precisely what one would expect for activated escape over a very high, relatively narrow potential barrier in the Kramers picture.

\begin{figure}[t]
 \centering
 \includegraphics[width=0.99\linewidth]{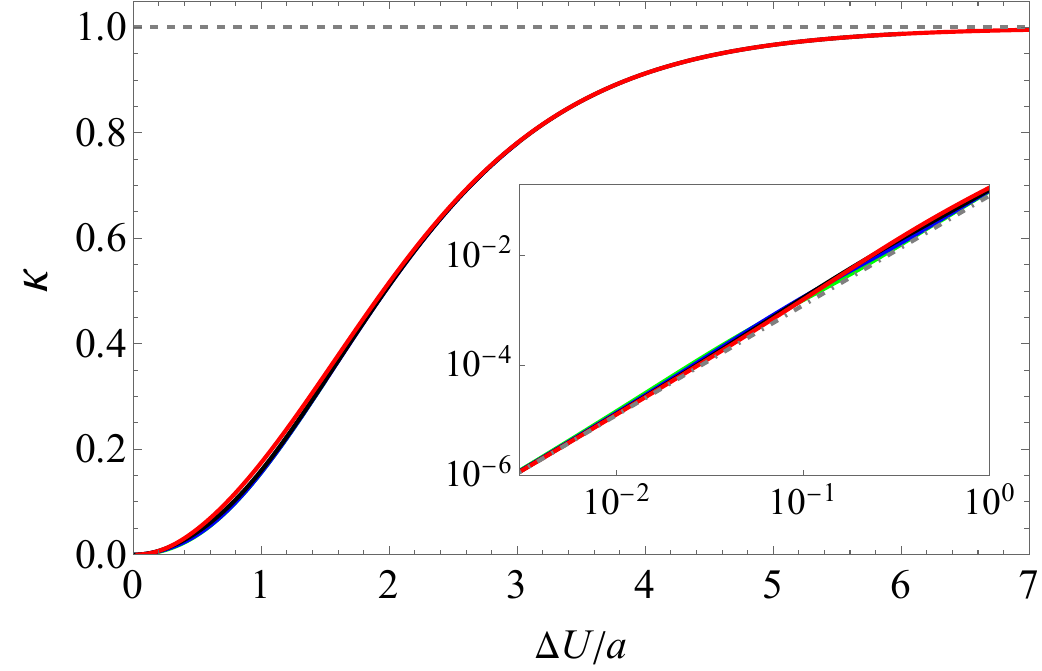}  
 \caption{
 The coefficient $\kappa$ in front of the Kramers factor $e^{-\Delta U}$, governing the decay rate of $P_n^*(y|x)$, plotted versus $\Delta U/a$ (see Eq.~\eqref{eq:exmplGenlongnSol}). Numerical solutions of Eq.~\eqref{eq:exmpldetzero} are shown for $a=3$ (red), $5$ (black), $10$ (blue), and $20$ (green). For $a\ge5$, the curves coincide. The dashed line represents the large-$\Delta U/a$ limit (Eq.~\eqref{eq:exmplpnlargeU}). The inset uses a log-log scale for the small-$\Delta U/a$ regime, where the dash-dot line indicates the theoretical limit (Eq.~\eqref{eq:exmpllargeApn}).
}\label{fig:expdecayKappa}
\end{figure}

{\it Case (ii): $a \to \infty$.}
Alternatively, if the barrier is made \emph{very wide} by taking $a\to\infty$ first (even if $\Delta U$ is large), Eq.~\eqref{eq:exmpldetzero} reduces to
\begin{equation}
    \label{eq:exmplLargeAlimit}
w\!\bigl(\tfrac{\Delta U}{a}, a\bigr)
\;\longrightarrow\;
\tfrac{1}{8}\,
\Bigl(\tfrac{\Delta U}{a}\Bigr)^2
\,c(\Delta U)\,
e^{-\Delta U}
\quad
\text{as}
\quad
a\to\infty
.
\end{equation}
Here $c(\Delta U)$ is determined by $\exp\left(-\Delta U\sqrt{1-c\exp(-\Delta U)}\right)=c\exp(-\Delta U)$ and smoothly approaches $1$ for both large and small $\Delta U$. 
Consequently, when $\Delta U/2a \to 0$, the dominant exponential decay in the propagator becomes
\begin{equation}
    \label{eq:exmpllargeApn}
P_n^*(y|x)
\;\sim\;
\exp\!\Bigl(
-\,\tfrac{1}{8}\,
\Bigl(\tfrac{\Delta U}{a}\Bigr)^2\,e^{-\Delta U}\,n
\Bigr)
\quad
\text{as}
\quad
\frac{\Delta U}{a}\to 0,
\end{equation}
meaning that the escape slows down further compared to the narrow-barrier scenario, owing to the extra factor $\bigl(\Delta U/a\bigr)^2$ in the exponent.

\begin{figure}[t]
 \centering
 \includegraphics[width=0.99\linewidth]{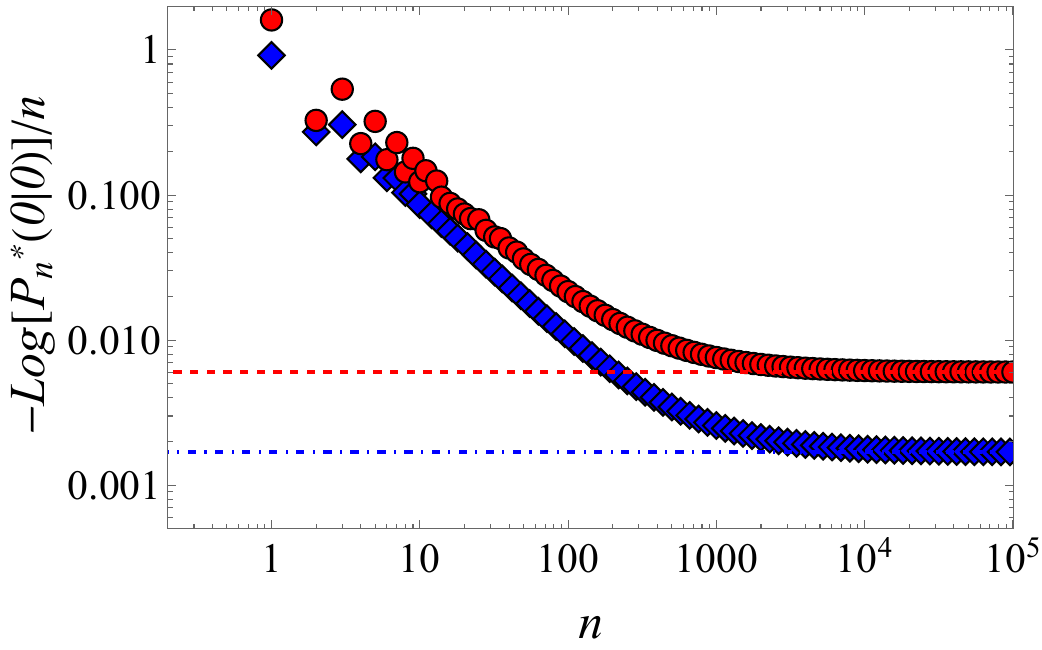}  
 \caption{
 Comparison of the exponential decay rate of the propagator for a random walker in an external potential (see Fig.\ref{fig:zigzagpotential}). Symbols represent results of numerical solution for $P_n^*(0|0)$, while the solid line corresponds to the prediction from $w$ in Eq.\eqref{eq:exmpldetzero}. Two cases are shown: (a) $a=5$, $p=0.3$ ({\color{blue}$\Diamond$}) and (b) $a=4$, $p=0.4$ ({\color{red}$\bigcirc$}). In both cases, the theoretical large‑$n$ limit of -$\log\bigl(P_n^*(0|0)\bigr)/n$ is determined from $w$. 
}\label{fig:numericconvergence}
\end{figure}

{\it Combined perspective.}
Overall, for large $n$,
\begin{equation}
\label{eq:exmplGenlongnSol}
P_n^*(y|x)\;\sim\;\exp\!\bigl(-\,\kappa\,e^{-\Delta U}\,n\bigr),
\end{equation}
with $\kappa$ dependent on both $\Delta U$ and $a$. In particular, when $a$ is large, $\kappa$ effectively becomes a function of the ratio $\Delta U/a$. 
In Fig.~\ref{fig:expdecayKappa}, we present the behavior of $\kappa$ by numerically solving Eq.~\eqref{eq:exmpldetzero}. Only small differences between the case $a=3$ and the cases where $a=5$, $a=10$, $a=20$, exist, indicating that beyond quite a small threshold, further widening the barrier barely changes the behavior of $\kappa$. 
The two different limiting behaviors derived above: 
$\kappa\to 1$  
when $
\frac{\Delta U}{a}\to\infty$ (Eq.~\eqref{eq:exmplLargeUlimit}),
and $\kappa\to \frac{1}{8}\,\bigl(\frac{\Delta U}{a}\bigr)^2$ 
when $\frac{\Delta U}{a}\to 0$ (Eq~\eqref{eq:exmplLargeAlimit})are displayed by the dashed line (main Fig.~\ref{fig:expdecayKappa}) and the dashed-dotted line (the inset).
Hence, the \emph{order} in which $\Delta U$ and $a$ become large is crucial for determining whether one observes the classic Kramers escape scenario or a significantly slower, wide-barrier decay rate.
In Fig.~\ref{fig:numericconvergence} we present excellent agreement between the developed theoretical results for the decay rate and the result of the numerical simulation of a process in the ``zigzag" potential. The convergence of $-\log\left(P_n^*(0|0)\right)/n$ in large $n$ limit to the theoretical result derived from Eq.~\eqref{eq:exmpldetzero} is evident.

The continuum limit of motion in piecewise linear potentials has been previously studied, specifically for a bistable linear potential~\cite{frisch1990exact} and a linear potential featuring a metastable state similar to our zigzag potential~\cite{privman1991exact}.
Unlike our present work, these earlier studies described motion via a Fokker–Planck equation solved in Laplace space by matching solutions across linear segments of the potential. Although these solutions contained the Kramers escape rate, the existence of two distinct asymptotic regimes revealed by our analysis (see the limiting behavior of $\kappa$ in Fig.\ref{fig:expdecayKappa} as $\Delta U\to\infty$) is absent.
We identified these distinct regimes explicitly due to our solution structure—Eq.\eqref{eq:gtildeZfinal form}—leading directly to Eq.~\eqref{eq:exmpldetzero}. Extending our discrete methodology to the continuous scenario to investigate analogous limits presents an interesting direction for future research.


\section{Discussion}
\label{sec:discussion}

Although research on random walks on lattices started a while ago~\cite{hughes1996random}, this field remains vibrant, driven by numerous emerging applications and novel analytical methods~\cite{grebenkov2024target,marris2023exact}. Recent applications span a variety of contexts, including restart and first-passage processes~\cite{bonomo2021first}, target search problems~\cite{grebenkov2024target}, dynamics on networks~\cite{schieber2023diffusion,riascos2021mean}, properties of hitting and splitting probabilities~\cite{levernier2021universality,klinger2022splitting}, exploration dynamics~\cite{regnier2023universal}, effects of disorder~\cite{shafir2024driven}, trapping phenomena~\cite{shafir2024disorder}, properties of rare events~\cite{vezzani2024fast} and interactions among multiple random walkers~\cite{barbier2022self}.

In this work, we have introduced an analytical approach aimed at addressing such diverse applications. Specifically, we established a mapping that enables leveraging insights from dynamics in simple potentials to characterize processes evolving under more complex potentials. This mapping is based on coupling two known methods: the Doob 
$h$-transformation~\cite{doob1984classical} and Montroll's defect theory~\cite{montroll1964random}. 
Individually, each method has inherent limitations and restrictive conditions; however, the synergy achieved through their combination provides a robust framework for analytically calculating propagators of random walks influenced by external potentials.

An important consequence of our method is evident from the structure of Eq.\eqref{eq:gtildeZfinal form}, which readily allows identification of additional zeros in z-space. 
Specifically, solving the equation $\det\left(\mathbf{I}-z\mathbf{M}\right)=0$ reveals new energetic states (corresponding to exponential decay rates) that were not present in the original, simpler scenario. 
We illustrated the usefulness of this approach by explicitly computing the exponential decay rate for a particle driven by a constant force in a potential with a metastable state. 
In particular, we identified two distinct regimes of the exponential decay rate in the limit of large $\Delta U$. 
Applying our developed method to other scenarios involving energetic barriers could offer further insights into the problem of optimization of Brownian escape by barrier reshaping~\cite{palyulin2012finite,chupeau2020optimizing}.

A natural next step is the generalization of our approach to stochastic processes evolving continuously in both space and time. 
Such an extension would significantly broaden the applicability of the framework introduced here.
Additionally, explicitly incorporating various boundary conditions, such as reflecting or absorbing boundaries, may reveal further contexts in which our method excels. 
Finally, we anticipate that our approach will be valuable in exploring the recently discovered phenomenon of Laplace universality~\cite{barkai2020packets,wang2020large}, particularly in scenarios involving binding external potentials. 

\section*{Acknowledgments}

I thank Shir Rajuan (Schondorf) for the helpful discussions that initiated this work.

\bibliography{synergyoftwo.bib}

\end{document}